\documentclass[a4paper,fleqn,usenatbib,useAMS]{mnras}

\usepackage{graphicx}	
\usepackage{amsmath}	
\usepackage{amssymb}	
\usepackage{multicol}        
\usepackage{multirow} 
\usepackage{bm}		
\usepackage{pdflscape}	
\usepackage{natbib}
\usepackage{times}
\usepackage{calc}
\usepackage{rotating}
\usepackage{lscape}
\bibpunct{(}{)}{;}{a}{}{,} 
\usepackage{longtable}
\usepackage{enumitem} 
\usepackage{hyperref}
\usepackage{pifont}
\usepackage{tikz}
\usepackage{capt-of}

\def\oversim#1#2{\lower0.5pt\vbox{\baselineskip0pt \lineskip-0.5pt
     \ialign{$\mathsurround0pt #1\hfil##\hfil$\crcr#2\crcr\sim\crcr}}}

\usepackage[T1]{fontenc}
\usepackage{ae,aecompl}

\usepackage{mathptmx}

\def\aap {{A\&A}}

\def\apjl {{ApJ}}

\def\apjs {{ApJs}}

\title[IPHASX J193718.6+202102]{Detailed studies of IPHAS sources - I. The disrupted late bipolar IPHASX J193718.6+202102}

\author[L. Sabin et al.] 
{L. Sabin$^{1}$\thanks{E-mail: lsabin@astro.unam.mx}, M.A.\,Guerrero$^{2}$, S. Zavala$^{3}$, J.A.\,Toal\'{a}$^{4}$, G.\,Ramos-Larios$^{5}$ and V. G\'omez-Llanos$^{1}$ \\  
$^{1}$Instituto de Astronom\'{\i}a, Universidad Nacional Aut\'onoma de M\'exico, Apdo. Postal 877, 22800 Ensenada, B.C., Mexico\\
$^{2}$Instituto de Astrof\'{i}sica de Andaluc\'{i}a (IAA-CSIC), Glorieta de la Astronom\'{i}a S/N, 18008 Granada, Spain\\
$^{3}$Tecnol\'ogico Nacional de M\'exico / I. T. Ensenada (TecNM/ITE), Blvd. Tecnol\'ogico No. 150, C. P. 22780, Ensenada, B. C., Mexico\\
$^{4}$Instituto de Radioastronom\'{i}a y Astrof\'{i}sica (IRyA), UNAM Campus Morelia, Apartado postal 3-72, 58090 Morelia, Michoacan, Mexico\\
$^{5}$Instituto de Astronom\'{i}a y Meteorolog\'{i}a, Universidad de Guadalajara, Av. Vallarta 2602, Arcos Vallarta, 44130 Guadalajara, Mexico\\
}

\pubyear{2020}

\begin{document}
\label{firstpage}
\pagerange{\pageref{firstpage}--\pageref{lastpage}}
\maketitle


\begin{abstract}

 
We present a detailed analysis of the new planetary nebula (PN) 
IPHASX J193718.6+202102 using deep imaging and 
intermediate- and high resolution spectroscopy that are interpreted 
through morpho-kinematic and photoionisation modelling. 
The physical structure of the nebula consists of a fragmented torus and an extremely faint orthogonal bipolar outflow, contrary to the pinched waist PN morphology suggested by its optical image. Our kinematic analysis indicates that the torus is expanding at 25$\pm$5 km s$^{-1}$ and is gradually breaking up. 
At an estimated distance of 7.1$_{-0.3}^{+0.8}$ kpc, the corresponding kinematic 
age of $\sim$26000 years is consistent with a faint and disintegrating PN. 
The intermediate-resolution spectra reveal  an excited PN with chemical 
abundances typical of Type II PNe.  
 Based on the latter we also estimate an initial mass for the progenitor in the range 
2--3~M$_{\sun}$ and a central star (CSPN) mass $M_\mathrm{CSPN}\sim$0.61~M$_{\sun}$.
The \emph{Spitzer} MIPS 24~$\mu$m emission that closely follows the fragmented 
torus  could be attributed to the emission of [O\,{\sc iv}] at 25.9 $\mu$m rather than to 
dust emission. 
All the results coherently point towards an evolved  moderately massive bipolar Type~II PN on the brink of dissolving into the interstellar medium.  
\end{abstract}

\begin{keywords}
ISM: kinematics and dynamics;  ISM: abundances; planetary nebulae: individual: IPHASX J193718.6+202102
  
\end{keywords}

\section{Introduction}\label{intro}
 
 The Isaac Newton Telescope (INT) Photometric H$\alpha$ survey \citep[IPHAS;][]{Drew2005,Gonzalez2008,Barentsen2014} was designed to provide a more complete sampling of the stellar population in the Northern Galactic Plane between $\pm$5$^{\circ}$. In particular, the survey has allowed the discovery of hundreds of new planetary nebulae (PNe), most of which either present a low surface brightness, which hampered their earlier detection, or were too compact to be clearly identified towards this crowded region of the Galactic Plane. 

These nebulae, which are ionized by the radiation field of evolved intermediate-mass stars, can be used to probe stellar mass loss and Galactic chemical evolution for example. 
IPHAS is an excellent tool to study very compact PNe \citep[which can be young and/or distant;][]{Viironen2009,Akras2019} as well as extended faint (and likely evolved) PNe \citep{Sabin2014}.
One of the most striking aspects of PNe is their morphology \citep{Balick2002,Sahai2011}. Indeed, most of them depart from spherical symmetry and the reason(s) for such behaviour has led to important progresses in research fields such as the study of binary systems \citep{Hillwig2016,Jones2017} and magnetic fields \citep{Asencio2014,Tocknell2014}. 
Advances were also made with kinematic studies and modelling \citep{Lopez2012,Sabin2017,GarciaDiaz2018}, chemical analyses \citep{Corradi2015,Jones2016,Wesson2018} and (magneto)hydrodynamics modelling \citep{Huarte2012,Balick2018}. 
These tools can be used to investigate the characteristics of the newly discovered IPHAS PNe which can differ from those of the generally brighter known PNe, hence providing a more global view of the PN population in the Galactic Plane. 
We therefore initiate here a series of publications dedicated to detailed 
analyses of individual IPHAS low surface brightness sources and PNe in 
particular.

In this article we present the case of the PN IPHASX J193718.6+202102 
(PN\,G056.1$-$00.4).
This object is included in the catalogue of\, \citet{Sabin2014} as a true PN, 
with the number 99 in its table~3, based on a follow up identification spectrum obtained at the Isaac Newton telescope (INT, La Palma). 
Accordingly, we will refer to it as Sab\,99 afterwards. 

The IPHAS images could only provide limited information on its morphology 
due to the short 120 seconds exposure in the wide H$\alpha$ filter, but they 
hint at a peculiar morphology. 
The nebula presents two parallel string-like features with discontinued emission resembling a set of knots. 
Such morphology led us to propose a classification as a bipolar PN with 
multiple shells or external structures and resolved internal structures 
(''Bas'', according to the classification scheme by \citealt{Sabin2014}).

In the following we present a more detailed analysis of Sab\,99 using new 
deeper observations. 
Imaging, intermediate- and high-dispersion spectroscopic observations are 
described in section \S\ref{obs}. 
The results of each method are shown in \S\ref{anal}. An infrared analysis based on \emph{Spitzer} data is also presented in section \S\ref{archive}. All the data will be used to establish a morpho-kinematic model to help us understand the evolutionary status of this PN (\S\ref{model}). 
Our discussion and concluding remarks are presented in \S\ref{summary} 
and \S\ref{Con}, respectively.


\section[]{Observations}\label{obs}

\subsection[]{NOT ALFOSC imaging}\label{obsim}

\begin{figure*}
\begin{center}
\includegraphics[width=2.1\columnwidth]{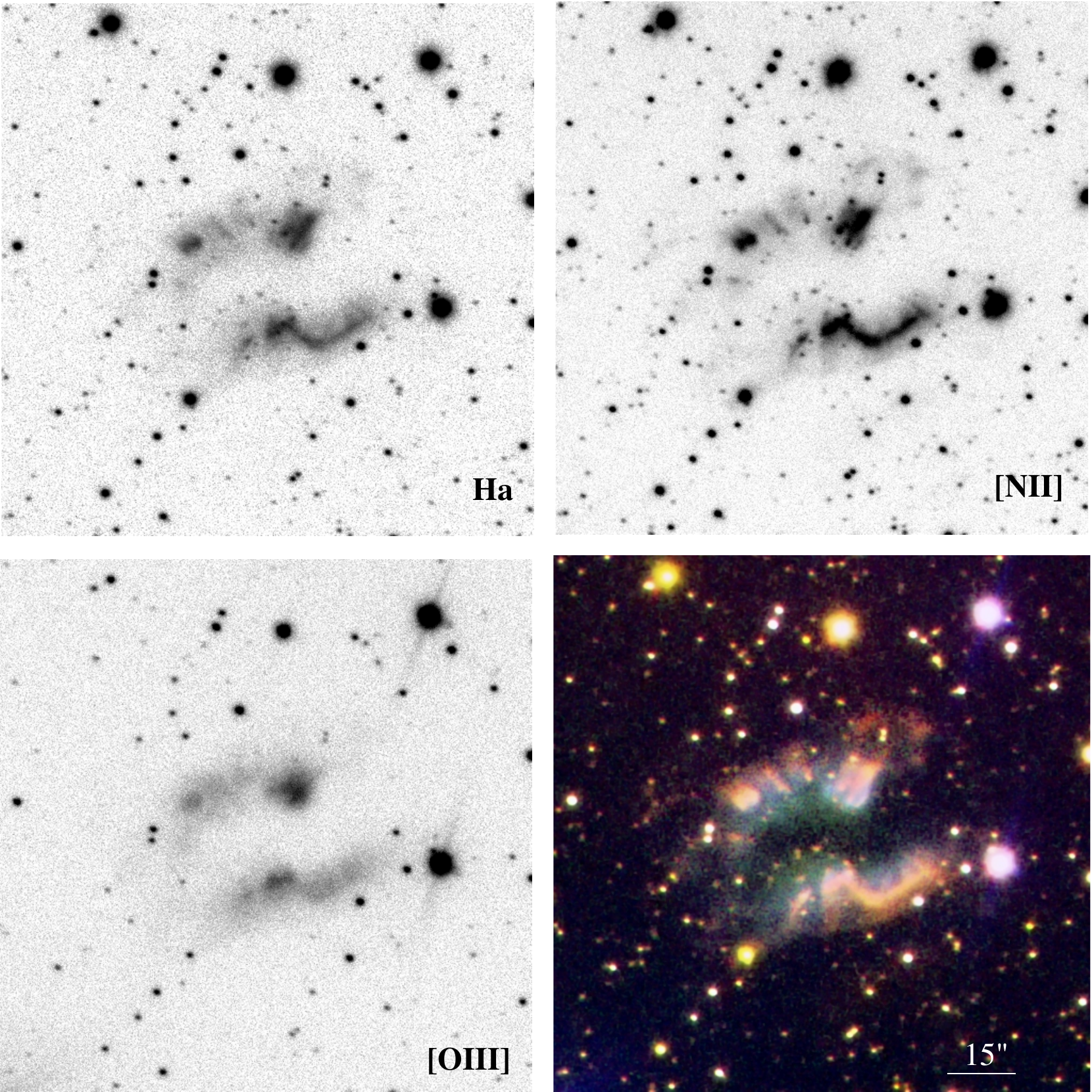}
\caption{
NOT ALFOSC narrow-band images of Sab\,99 and color composite picture with 
[N\,{\sc ii}] in red,  H$\alpha$ in green, and [O\,{\sc iii}] in blue. 
North is up, East to the left. 
}
\label{NOT}
\end{center}
\end{figure*}

Sab\,99 was observed with the Alhambra Faint Object Spectrograph and Camera (ALFOSC) mounted on the 2.5~m Nordic Optical Telescope (NOT) at the Observatorio Roque de los Muchachos (ORM, La Palma, Spain). The camera is equipped with a 2048$\times$2048 EEV CCD (13.5\,$\mu$m pixel size) detector providing a plate scale of 0$\farcs$184 pix$^{-1}$ and a field of view of 6$\farcm$3$\times$6$\farcm$3. 
Three sets of narrow-band images with exposure times 3$\times$600s were 
obtained on 2016 November 27 using the 
H$\alpha$ (\mbox{$\lambda_c = 6567$ \AA}, \mbox{FWHM $= 8$ \AA )}, 
[N\,{\sc ii}] (\mbox{$\lambda_c = 6588$ \AA}, \mbox{FWHM $= 9$ \AA)}, and 
[O\,{\sc iii}] (\mbox{$\lambda_c = 5007$ \AA}, \mbox{FWHM $= 30$ \AA)} filters. 
The observing conditions were excellent with a seeing between  0$\farcs$6 and 0$\farcs$8 based on the FWHM of stars in the field. 
All the images were reduced using standard {\sc iraf}\footnote{
IRAF is distributed by the National Optical Astronomy Observatories, which is operated 
by the Association of Universities for Research in Astronomy, Inc. (AURA) under 
cooperative agreement with the National Science Foundation. 
} 
routines \citep{Tody1986}.
The individual narrow-band images and a color-composite picture are 
presented in Figure~\ref{NOT}.

\begin{figure}
\begin{center}
\hspace*{-1.4cm}
\includegraphics[height=8.4cm]{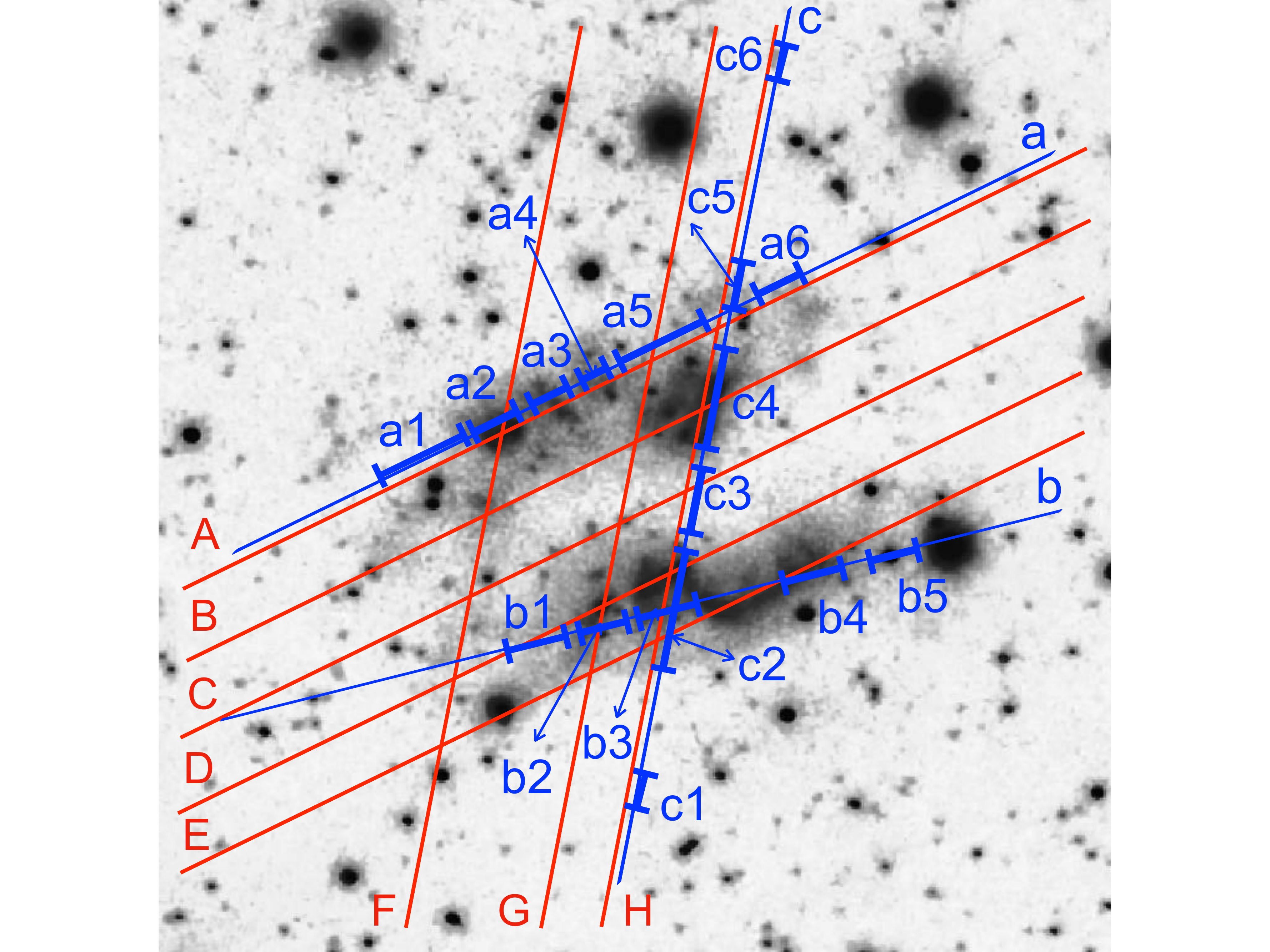}
\caption{
MES (red, upper-case letters) and GTC OSIRIS (blue, lower-case letters) slit 
positions superimposed on an image of Sab\,99. 
The GTC OSIRIS slits {\it a} and {\it c} are spatially coincident with the MES 
slits A and H, respectively, but were displaced for clarity. 
The regions of the GTC OSIRIS spectra used for chemical analysis are 
labelled on the image as a1--a6, b1--b5, and c1--c6. 
}
\label{slit}
\end{center}
\end{figure}

\subsection[]{MES high resolution spectroscopy}\label{obsmes}

Long-slit high dispersion optical spectra of Sab\,99 were acquired with the
Manchester Echelle Spectrograph (MES, \citealt{Meaburn2003}) installed on the
2.12~m telescope at the Observatorio Astron\'omico Nacional, San Pedro
M\'artir (OAN-SPM, Mexico). 
The detector used was a 2048$\times$2048 pixels E2V CCD with a pixel size of
13.5 $\mu$m.  
The slit length was 6$\farcm$5 and the slit width was 150 $\mu$m corresponding 
to 1$\farcs$9. 
The observations were obtained with 2$\times$2 and 4$\times$4 binning, 
resulting in spatial scales and spectral dispersions of 0$\farcs$351 
pix$^{-1}$ and 0.05 \AA\,pix$^{-1}$, and 0$\farcs$702 pix$^{-1}$ 
and 0.1 \AA\,pix$^{-1}$, respectively. 
An H$\alpha$ filter with $\Delta\lambda$ = 90 \AA\ was used to isolate 
the 87th echelle order including the H$\alpha$ and the [N~{\sc ii}] 
$\lambda\lambda$6548,6584 lines. 
Eight 1800s spectra were obtained with the slits placed across different 
nebular features to provide a complete map the velocity field of Sab\,99. 
Table~\ref{Mezobs} summarizes the different observations realized with 
MES and the position of each slit is shown in Figure~\ref{slit}. 
A ThAr lamp was used for wavelength calibration and the 
data were reduced using standard {\sc iraf} packages. 

The FWHM of the arc lamp emission lines indicate a spectral resolution  $\sim12$ km\,s$^{-1}$.

\begin{table}
\begin{center}
\caption{\label{Mezobs} MES Observing log}
\begin{tabular}{|c|c|c|c|c|}
\hline 
Slit & Observation Date & PA           & Binning  & Seeing   \\
     &                  & ($^{\circ}$) & pixels   & ($\arcsec$) \\
\hline 
A & 2017/08/05 & $-$65 & 2$\times$2   & 1.4    \\ 
B & 2018/05/27 & $-$65  & 2$\times$2  & 1.2   \\ 
C & 2018/05/27 & $-$65  & 4$\times$4  & 1.3   \\ 
D & 2017/08/06 & $-$65& 2$\times$2   &  1.3   \\ 
E & 2017/08/06 & $-$65& 2$\times$2   &  1.2  \\ 
F& 2018/05/28 & $-$10 & 2$\times$2   &  1.3   \\ 
G &2018/05/27 & $-$10 & 2$\times$2   & 1.2     \\ 
H & 2017/08/06 & $-$10& 2$\times$2   & 1.4  \\
\hline
\end{tabular} 
\end{center}
\end{table}

\subsection[]{GTC OSIRIS intermediate resolution spectroscopy}\label{obsgtc}

Intermediate resolution spectroscopy was performed with the 10.4~m Gran Telescopio Canarias 
(GTC) and the Optical System for Imaging and low-Intermediate-Resolution Integrated Spectroscopy 
OSIRIS at ORM (La Palma, Spain). Two Marconi CCD42-82 (2048$\times$4096 pixels) detectors 
were used and a 2$\times$2 binning was employed, leading to a spatial scale of 0$\farcs$254 pix$^{-1}$. 
The R1000B grism was used, providing a spectral coverage 3630-7500 \AA\ 
and a dispersion of 2.12 \AA\ pix$^{-1}$. 
In order to obtain a good coverage of the nebula, spectra were obtained 
for three slit positions across morphological features of interest (see
Tab.~\ref{GTCobs} and Fig.~\ref{slit}). 
In all cases the slit length was 7$\farcm$4 and the slit width 0$\farcs$8, leading to  a spectral resolution of 6.15~\AA.

The exposure time was 3$\times$1200~s for each slit. 
The wavelength calibration was performed with HgAr and Ne lamps and the 
flux calibration was achieved with the standard stars GD140, GD190 and 
Feige\,110. The data reduction was carried out using standard {\sc iraf} routines.


\begin{table}
\begin{center}
\caption{\label{GTCobs} GTC Observing log}
\begin{tabular}{|c|c|c|c|}
\hline 
Slit & Observation Date & PA & Seeing \\ 
 & & ($^{\circ}$) & ($\arcsec$) \\
\hline 
a & 2018/05/22 & $-$65 & 1.0 \\ 
b & 2018/06/04 & $-$78 & 0.8 \\ 
c & 2017/08/19 & $-$10 & 1.3 \\ 
\hline 
\end{tabular} 
\end{center}
\end{table}

\subsection[]{OSN broad band imaging}\label{obsgtc}

Broadband $U$, $B$, $V$ and $R$ images were obtained on 2019 November 7 
with the 1.5m telescope located at the Sierra Nevada observatory 
(OSN, Spain) as part of a  Director Discretionary Time program.  
We used the 2048$\times$2048 pixels CCDT150 camera with a pixel size of 
13.5$\mu$m, resulting in a plate scale of 0$\farcs$232 pix$^{-1}$ and a 
field of view of 7$\farcm$92$\times$7$\farcs$92.  
A 2$\times$2 binning factor was employed.  
In total, we obtained 6 exposures of 100s in each filter except the U 
filter where we used 6 exposures of 600s. 
The photometric calibration was performed with the Landolt fields G26-7 
and G93-48.  	

\subsection[]{{\it Spitzer} MIPS imaging}
\label{spitzer}

We used the NASA/IPAC Infrared Science Archive (IRSA) and retrieved \emph{Spitzer}/MIPS (Multiband Imaging Photometer for  \emph{Spitzer}, \citealt{Rieke2004}) data from the MIPS Galactic Plane legacy survey MIPSGAL \footnote{https://irsa.ipac.caltech.edu/data/SPITZER/MIPSGAL/}. MIPS, which has a 128$\times$128 Si:As detector and a 24$\mu$m pixel size, provides image and photometric data in the bands centered at 24, 70, and 160 $\mu$m  with spatial resolution of 6$\farcs$0, 18$\farcs$0, and 40$\farcs$0, respectively.


\begin{figure*}
\begin{center}
\includegraphics[width=1\linewidth]{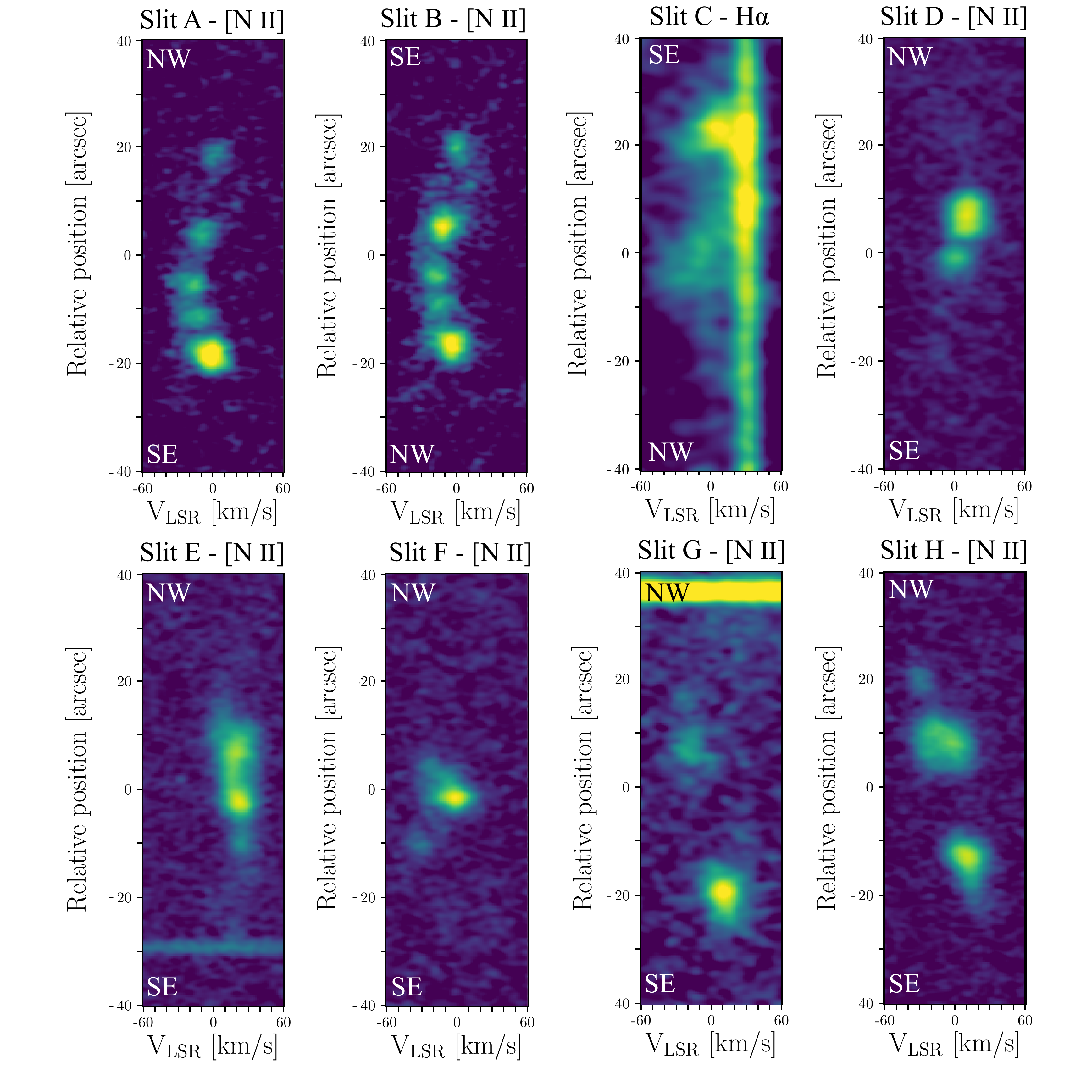}
\vspace*{-0.6cm}
\caption{
MES PV diagrams of the [N\,{\sc ii}] $\lambda$6583 \AA\ 
emission line, except for slit C, which shows the H$\alpha$ emission line.  
The bright line in the PV diagram of slit C at velocity $\sim$30 km~s$^{-1}$ 
is a telluric line, which was not removed to avoid affecting the spectrum. 
All the images were Gaussian smoothed to enhance faint structures. 
}
\label{Kine1}
\end{center}
\end{figure*}

\section[]{Analysis}\label{anal}

\subsection[]{Morphology}\label{im}

The high quality NOT imagery presented in Figure~\ref{NOT} reveals 
with great detail the morphology of Sab\,99. 
The brightest components of Sab\,99 are two $\sim$1$\farcm$0$\times$0$\farcm$7 
filamentary string-like features that mirror each other's curvature. 
They present a fragmented morphology, which is best seen in the [N\,{\sc ii}] 
light, where bright knots are followed by tails reminiscent of ionization 
shadows. 
Other PNe such as the faint K\,1-10 \citep{Corradi1995}, PN\,G321.6+02.2 
\citep{Corradi1997}, and the bright NGC\,2818 \citep{Vazquez2012} show 
similar symmetric rims.  
In those cases these opposite symmetric arms draw the walls and waist of 
bipolar outflows.

The NOT images do not reveal any nebulosity surrounding the main bright 
central structure. 
It was not possible neither to identify a central star (CS) based on the 
UV-Excess Survey of the Northern Galactic Plane \citep[UVEX,][]{Groot2009}. 
The observations obtained at the OSN did not detect a clear central star 
up to a limiting magnitude of 20~mag in the RGO $U$ filter of UVEX.



\subsection[]{Kinematics}\label{mezcal}

The eight position-velocity (PV) diagrams shown in Figure~\ref{Kine1} give 
a relatively complete view of the velocity field distribution in Sab\,99. 
Slits A, B, D and E, going across the opposite arms  (see their location 
in Fig.~\ref{slit}), underline the fragmented nature of the [N\,{\sc ii}]~$\lambda$6583~\AA\ emission. 
The emission observed in these PV diagrams seems to show a faint curved expansion 
below and above the bright central region. However, these patterns are not an indication 
for the presence of additional emission from an extended outflow along the slit, as 
the full extent $\sim$50$\arcsec$ of the [N\,{\sc ii}] emission in these diagrams 
is consistent with the size of the nebula in the narrow-band images in 
Figure~\ref{NOT}. 
The H$\alpha$ emission is concentrated in the center of the nebula as shown by the 
slit C and tends to show some curvature.

\begin{figure}
\begin{center}
\includegraphics[height=8.4cm]{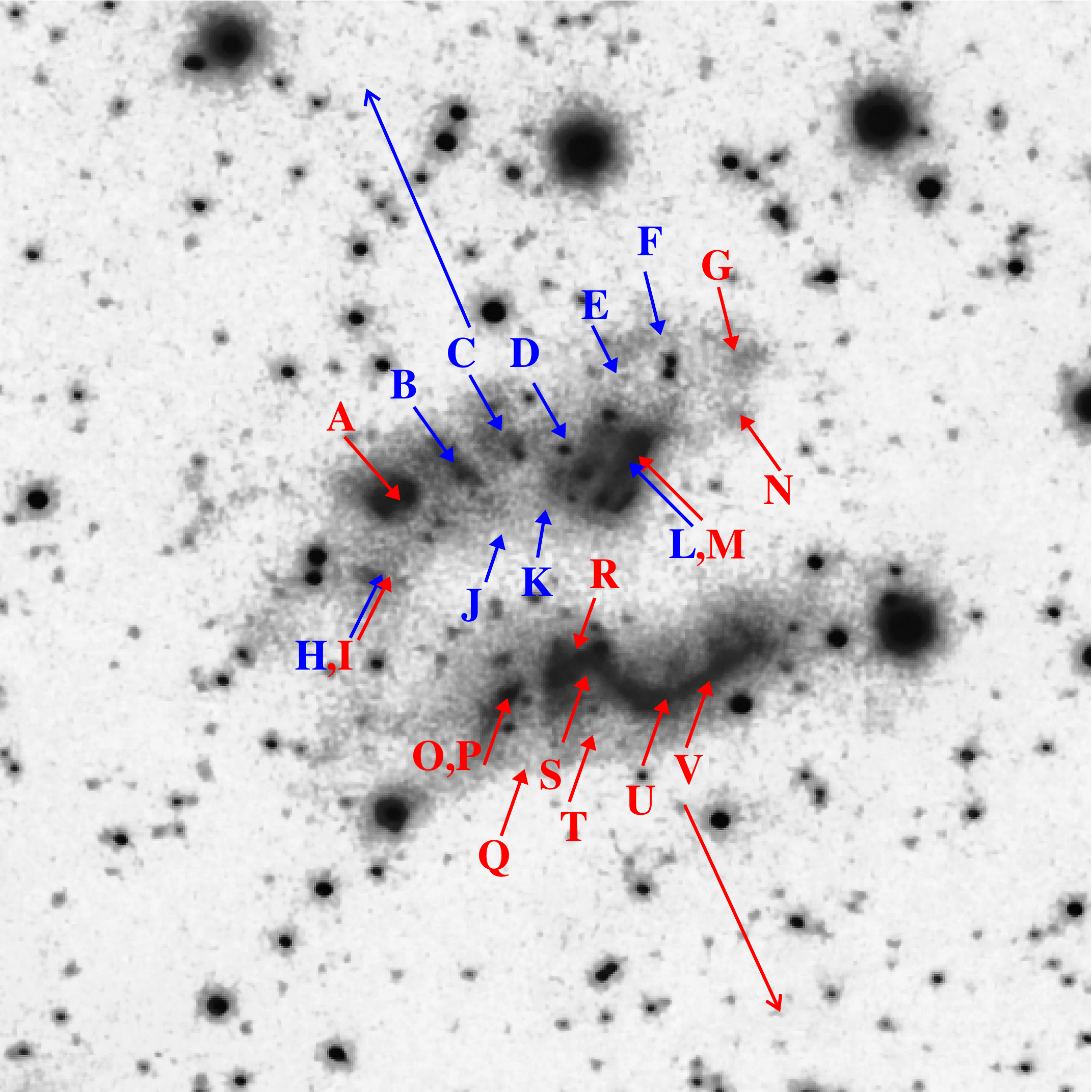}
\caption{
Spatial distribution on the different knots in Sab\,99 for which their radial 
velocities have been measured (see Table \ref{Kine3}). 
The blue arrows and letters illustrate the generally blue-shifted motion of the 
Northern arm and the red arrows and letters the global red-shifted motion of the 
Southern arm. 
}
\label{Kine2}
\end{center}
\end{figure}

Since it was not possible to derive the systemic velocity of the PN due to 
its morphology and the impossibility to precisely locate the central star, the radial velocities of individual features in the [N\,{\sc ii}] 
PV diagrams in Table~\ref{Kine3} are only corrected from the Local Standard of 
Rest. 
The distribution of these velocities is illustrated in Figure~\ref{Kine2},  
with peak values ranging from $+$30.2 km~s$^{-1}$ to $-$27.8 km~s$^{-1}$. 
Apart from some outliers, a general trend can be observed with the upper arm being 
blue-shifted, while the lower one is red-sifted. 
The velocity distribution of both structures is not symmetric though, 
but we note that the FWHM for each data point is quite large, with a 
mean of 20.8 km~s$^{-1}$.

We also note that the velocity variations found among adjacent knots, such as the knots I, J and K sampled by Slit B, are indicative of the loose spatio-kinematic structure of the torus (see Section~6).

\begin{table}
\begin{center}
\caption{
\label{Kine3} 
Radial velocities (in the LSR) and FWHM of bright features covered 
by slits A, B, D, and E.  
Each feature can spatially be identified on Figure~\ref{Kine2}.
}
\begin{tabular}{|c|r|r|c|r|r}
\hline 
Feature & Velocity & FWHM & Feature & Velocity & FWHM \\
      & 
\multicolumn{2}{c}{(km s$^{-1}$)} &
      &
\multicolumn{2}{c}{(km s$^{-1}$)} \\
\hline
A &  $+$0.7~~~ & 20~~ & L &  $-$4.0~~~ & 20~~ \\
B & $-$10.8~~~ & 25~~ & M &  $+$9.5~~~ & 37~~ \\
C & $-$15.4~~~ & 25~~ & N &  $+$7.9~~~ & 14~~ \\
D & $-$10.7~~~ & 25~~ & O & $+$15.4~~~ & 19~~ \\
E &  $-$6.1~~~ & 22~~ & P & $+$16.1~~~ & 23~~ \\
F & $-$13.2~~~ & 14~~ & Q & $+$30.1~~~ & 13~~ \\
G &  $+$5.7~~~ & 18~~ & R & $+$24.7~~~ & 20~~ \\
H &  $+$2.4~~~ & 23~~ & S & $+$28.2~~~ & 22~~ \\
I & $-$27.8~~~ & 19~~ & T & $+$28.2~~~ & 17~~ \\
J &  $-$6.7~~~ & 24~~ & U & $+$24.2~~~ & 20~~ \\
K &  $-$9.6~~~ & 18~~ & V & $+$26.0~~~ & 21~~ \\
 \hline
\end{tabular}
\end{center}
\end{table}
\subsection[]{Chemistry}\label{gtc}
 
The chemical analysis of Sab\,99 was performed with the GTC OSIRIS data at three position angles going along both strings of knots (PA=$-$65$^{\circ}$ and $-$78$^{\circ}$) and across them (PA=$-$10$^{\circ}$), as illustrated 
in Fig.~\ref{slit}.  

One-dimensional spectra were extracted with the {\sc iraf} task {\it apall} from 17 different regions to map the chemistry across the nebula\footnote
 
The PA on the sky of slits {\it a} and {\it b} were close to the parallactic angle, but that of slit {\it c} was mostly orthogonal, $\simeq80^\circ$.  
At the airmass of the latter observation, $\simeq1.2$, the differential chromatic diffraction between the most extremes [Ne~{\sc iii}] $\lambda$3869 \AA\ and [Ar~{\sc iii}] $\lambda$7751 \AA\ emission lines can amount up to 0\farcs8, but it is $\leq0\farcs4$ for the most important lines in the range 4800--6800 \AA\ \citep{Filippenko1982}. 
The slit width of 0$\farcs$8 is just twice that value, but the diffuse nature of this source, with knots much more extended than the slit width and differential chromatic diffraction shift, minimizes the impact of the latter. 
This is further diluted by the use of aperture sizes significantly larger than the maximum value of the differential chromatic diffraction. 
.  
The line fluxes were determined with the {\sc iraf} task {\it splot} and the errors associated to each intensity takes into account the statistical error (noise around the position of the corresponding line) and the errors due to the flux calibration. The data analysis was performed using the code {\sc PyNeb} by \citet{Luridiana2015} 
and the data were corrected from extinction using the \citet{Cardelli1989} law with R$_{v}$=3.1. 
Uncertainties in the physical conditions and ionic abundances determined by {\sc PyNeb} were estimated using a Monte Carlo simulation of 1000 random values with a normal distribution around the observed intensity and a standard deviation ($\sigma$) of each line.
The results of the analysis are shown in Tables \ref{GTCA}, \ref{GTCB}, and \ref{GTCC}. We emphasize that we excluded throughout this analysis all measurements with errors greater than 50\% (which roughly corresponds to a signal to noise ratio lower than 3). As a consequence, some regions are not included in the aforementioned tables due to the large errors or absence of the Balmer lines.

Sab\,99 is characterised by a relatively large extinction and we derived mean logarithmic extinctions  using the \citet{Cardelli1989} extinction law of <c(H$\beta$)> = 2.47$\pm$0.23 along the upper arm (slit a), <c(H$\beta$)> = 2.59$\pm$0.18 along the lower arm (slit b), and 
<c(H$\beta$)> = 2.56$\pm$0.12 for the transverse region (slit c). The values are relatively consistent with each other and lead to a mean c(H$\beta$) of 2.54$\pm$0.18 for the whole nebula. \\

\begin{table}
\begin{center}
\caption{\label{GTCA} 
Intrinsic intensities of emission lines in regions of the upper arm.} 
\setlength{\tabcolsep}{0.35\tabcolsep}
\begin{tabular}{@{\extracolsep{4pt}}|l|c|c|c|c|c|@{}}
\hline 
Ion &  $\lambda_{0}$ & a2& a3 & a4 & a5 \\
\hline 
     H$\gamma$       &4340&            75$\pm$23  &  ...         &   ...       &  ...                   \\
     HeII            &4686&            39$\pm$7   & 53$\pm$14    &  49$\pm$11   &   ...                 \\
    H$\beta$         &4861&            100$\pm$12  & 100$\pm$17   &  100$\pm$18 & 100$\pm$38                    \\
    \,[O\,{\sc iii}] &4959&            447$\pm$45 & 435$\pm$60   &  463$\pm$67 & 626$\pm$175        \\
    \,[O\,{\sc iii}] &5007&            1306$\pm$129& 1432$\pm$191 & 1423$\pm$199& 1714$\pm$478      \\
   \,[N\,{\sc i}]    &5199&            9$\pm$3    &  ...         & ...         &    ...               \\
   \,[N\,{\sc ii}]   &5755&            7$\pm$2    & ...          &  ...        &  ...               \\
    HeI              &5876&            15$\pm$3   & 9$\pm$3      &  ...        &  ...               \\ 
    \,[O\,{\sc i}]   &6300&            35$\pm$5   & 15$\pm$3     &   8$\pm$2   &    ...               \\
    \,[O\,{\sc i}]   &6363&             8$\pm$1   & ...          &   ...       &  ...              \\ 
    \,[N\,{\sc ii}]  &6548&            183$\pm$25 & 88$\pm$17    &  68$\pm$13   & 63$\pm$25          \\
   H$\alpha$         &6563&           287$\pm$39  & 287$\pm$53   & 287$\pm$56  & 287$\pm$112        \\
    \,[N\,{\sc ii}]  &6583&            541$\pm$74 & 267$\pm$50   & 219$\pm$43  & 208$\pm$82     \\
    HeI              &6678&           3.4$\pm$0.7 & 2.8$\pm$1.2  &  ...        & ...                   \\
    \,[S\,{\sc ii}]  &6716&            50$\pm$7   & 38$\pm$8     &  36$\pm$8   & 36$\pm$15          \\
    \,[S\,{\sc ii}]  &6731&            39$\pm$6   & 27$\pm$5     &  26$\pm$6   & 23$\pm$10          \\
    HeI              &7065&            2.6$\pm$0.5    &  ...     &  2.3$\pm$0.9 &    ...            \\ 
  \,[Ar\,{\sc iii}]  &7136&           29$\pm$5    & 30$\pm$7     &  30$\pm$7   & 30$\pm$14           \\
   \,[O\,{\sc ii}]   &7320&            5.5$\pm$0.9&  ...         &  6$\pm$2    &    ...           \\
  \,[O\,{\sc ii}]    &7330&            3.8$\pm$0.7&  ...        &   ...        &  ...            \\
 \,[Ar\,{\sc iii}]   &7751&            4.9$\pm$0.9&  7$\pm$2     &   5$\pm$1   &  ...              \\
 \hline 
    c(H$\beta$)   &                &    2.40$\pm$0.14  &  2.43$\pm$0.19     &  2.40$\pm$0.20   & 2.63$\pm$0.40  \\
 \hline 
\end{tabular} 
\begin{minipage}[!t]{0.7\textwidth}
\end{minipage}
\end{center}
\end{table}


 \begin{table}
\begin{center}
\caption{\label{GTCB} 
Intrinsic intensities of emission lines in regions of the lower arm.}
\setlength{\tabcolsep}{0.35\tabcolsep}
\begin{tabular}{@{\extracolsep{4pt}}l|c|c|c|c|c|@{}}
\hline 
 Ion &  $\lambda_{0}$ &  b2 &b3 & b4 & b5\\
\hline 
     HeII            &4686&             105$\pm$25    & 72$\pm$11    & 67$\pm$10  & ... \\
    H$\beta$         &4861&             100$\pm$21   & 100$\pm$13   &100$\pm$14  &  100$\pm$16 \\
    \,[O\,{\sc iii}] &4959&             447$\pm$74   & 405$\pm$44  &395$\pm$47 & 511$\pm$68 \\
    \,[O\,{\sc iii}] &5007&             1400$\pm$226 & 1189$\pm$127 &1194$\pm$137&1481$\pm$193\\
   \,[N\,{\sc ii}]   &5755&             ...          & 4.0$\pm$0.9     &6$\pm$1    &...\\
    HeI              &5876&             ...          & 6$\pm$1     &7$\pm$1    & ...\\
    \,[O\,{\sc i}]   &6300&             18$\pm$5     & 22$\pm$3    &20$\pm$3   & 29$\pm$5\\
    \,[O\,{\sc i}]   &6363&             8$\pm$3      & 8$\pm$1     &5.3$\pm$1.3    &7$\pm$1 \\
    \,[N\,{\sc ii}]  &6548&             58$\pm$13     & 113$\pm$17   &101$\pm$16   &128$\pm$23\\
   H$\alpha$         &6563&             287$\pm$65   & 288$\pm$43  &286$\pm$46 & 287$\pm$52\\
    \,[N\,{\sc ii}]  &6583&             185$\pm$42   & 345$\pm$51  &319$\pm$51 &390$\pm$71\\
    HeI              &6678&             2.7$\pm$1.2      &  3$\pm$1    &2.8$\pm$0.7    &5$\pm$2 \\
    \,[S\,{\sc ii}]  &6716&             26$\pm$6     & 40$\pm$6    &36$\pm$6   &33$\pm$7\\
    \,[S\,{\sc ii}]  &6731&             22$\pm$5     & 35$\pm$5    &29$\pm$5   &23$\pm$5\\
    HeI              &7065&              ...         & ...         &1.6$\pm$0.4    & ...\\
  \,[Ar\,{\sc iii}]  &7136&             29$\pm$8     & 33$\pm$6    &33$\pm$6   &33$\pm$7 \\
   \,[O\,{\sc ii}]   &7320&              ...         & 6$\pm$1     &4.5$\pm$1.2    & ...\\
  \,[O\,{\sc ii}]    &7330 &             ...         & 3.1$\pm$0.7 &3.7$\pm$0.8    & ...\\
  \,[Ar\,{\sc iii}]   &7751&             3.2$\pm$1.0     & 3$\pm$1     &3.4$\pm$0.7    &...\\ 
 \hline 
  
  c(H$\beta$)   &  & 2.64$\pm$0.23  & 2.49$\pm$0.15  & 2.42$\pm$0.17  & 2.81$\pm$0.19  \\
 \hline 
\end{tabular} 
\begin{minipage}[!t]{0.6\textwidth}
\end{minipage}
\end{center}
\end{table}

\begin{table}
\begin{center}
\caption{\label{GTCC}  Extracted regions with dereddened (I/H$\beta$) fluxes across the PN or Slit c}.
\setlength{\tabcolsep}{0.35\tabcolsep}
\begin{tabular}{@{\extracolsep{4pt}}l|c|c|c|@{}}
\hline 
Ion &  $\lambda_{0}$ &  c2 & c4 \\
\hline 
   \,[Ne\,{\sc iii}] &3869 &                    ...        &                    217$\pm$55          \\ 
   \,[Ne\,{\sc iii}] &3968 &                    ...        &                  110$\pm$31          \\
    H$\gamma$        &4340 &               62$\pm$17       &                  77$\pm$14         \\
     HeII            &4686 &               72$\pm$8        &                  66$\pm$6          \\
    H$\beta$         &4861 &               100$\pm$10       &                100$\pm$9         \\
    \,[O\,{\sc iii}] &4959 &               418$\pm$38      &                  499$\pm$42          \\
    \,[O\,{\sc iii}] &5007 &               1251$\pm$111     &                 1464$\pm$123   \\
      HeI            &5411 &                 5$\pm$1       &                2.6$\pm$0.6          \\
     \,[N\,{\sc ii}] &5755 &                    ...        &                2.7$\pm$0.5             \\
    HeI              &5876 &                    ...        &                11$\pm$1           \\
    \,[O\,{\sc i}]   &6300 &               17$\pm$2        &                10$\pm$1           \\
\,[S\,{\sc iii}]+HeII&6312 &                    ...        &                3.3$\pm$0.4            \\
    \,[O\,{\sc i}]   &6363 &               4.5$\pm$0.7         &                3.2$\pm$0.4            \\
    \,[N\,{\sc ii}]  &6548 &               97$\pm$12      &                 86$\pm$10      \\
   H$\alpha$         &6563 &               287$\pm$36       &               290$\pm$34    \\
    \,[N\,{\sc ii}]  &6583 &               313$\pm$39       &               261$\pm$31   \\
    HeI              &6678 &                2.1$\pm$0.6        &                3.2$\pm$0.6             \\
    \,[S\,{\sc ii}]  &6716 &               39$\pm$5        &                35$\pm$4            \\
    \,[S\,{\sc ii}]  &6731 &               33$\pm$5        &                30$\pm$4            \\
    HeI              &7065 &                2.1$\pm$0.5        &                2.1$\pm$0.4             \\
  \,[Ar\,{\sc iii}]  &7136 &               30$\pm$4        &                33$\pm$4            \\
   \,[O\,{\sc ii}]   &7320 &               ...             &                4.0$\pm$0.8            \\
  \,[O\,{\sc ii}]    &7330 &               ...             &                2.0$\pm$0.3             \\
  \,[Ar\,{\sc iii}]  &7751 &               4.4$\pm$0.8         &                5$\pm$1             \\
 \hline 
    c(H$\beta$)   &  & 2.58$\pm$0.13  &2.55$\pm$0.12 \\
\hline  
\end{tabular} 
\begin{minipage}[!t]{0.80\textwidth}
\end{minipage}
\end{center}
\end{table}

\begin{figure}
\begin{center}
\includegraphics[width=\linewidth]{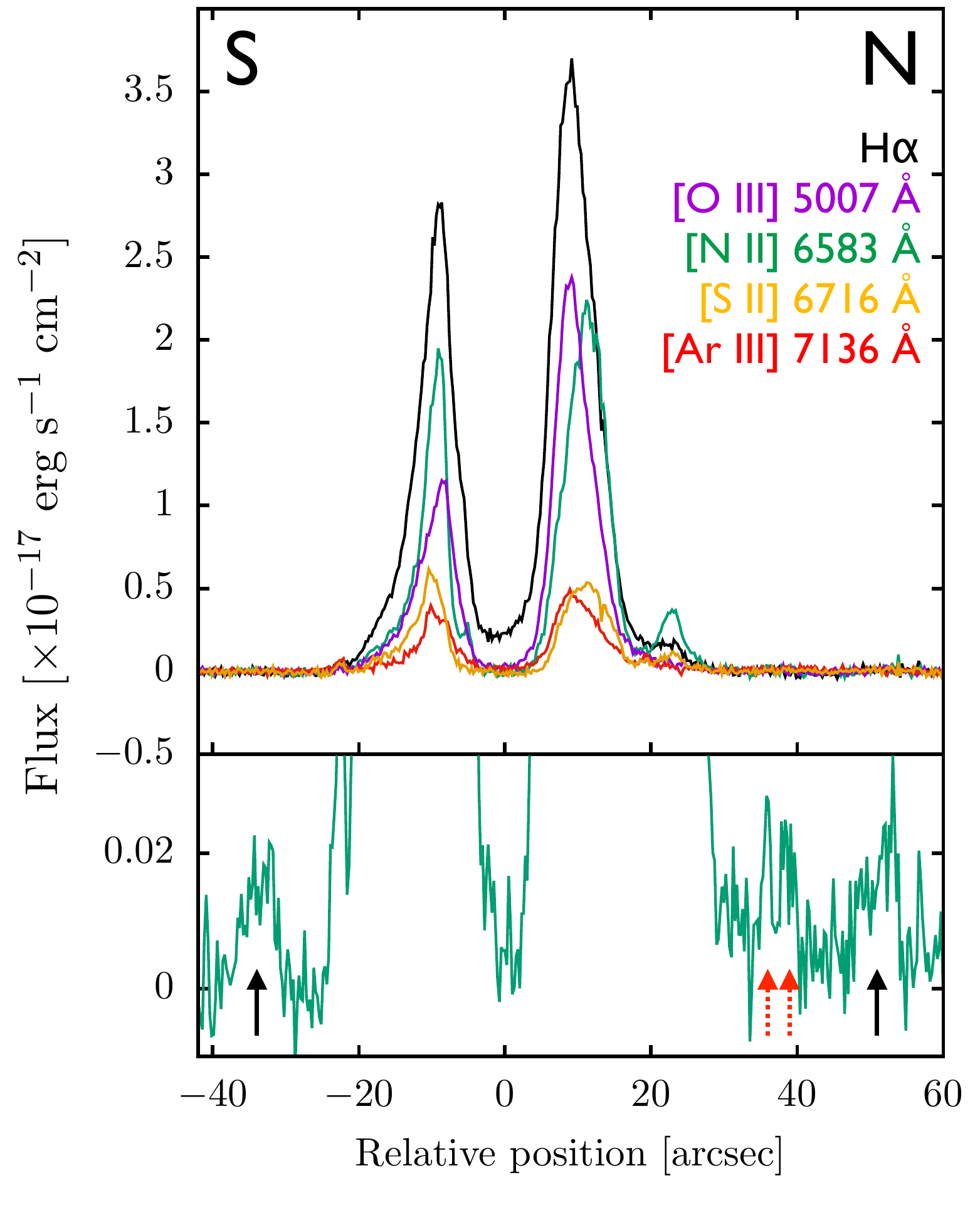}
\vspace*{-0.6cm}
\caption{Spatial profiles of various emission 
lines in Sab\,99 obtained through the GTC OSIRIS slit c. S and N show the southern and northern directions, respectively. 
The bottom panel presents a zoom of the [N~{\sc ii}] profile to show the position of the clumps c1 and c6 (see Figure~\ref{slit}) marked with black solid black arrows (the red dashed-line arrows mark the location of the continuum emission from background stars).}
\label{profiles}
\end{center}
\end{figure} 

 It was possible to derive the electronic temperatures and corresponding densities in 
the brightest a2, b3, b4 and c4 regions due to the presence of the \,[N\,{\sc ii}]\,5755~\AA\, line and sulfur doublet [S\,{\sc ii}]~6716,6731~\AA. 
The mean values obtained from the Monte Carlo distribution, as well as the associated standard deviations, are shown in Table \ref{TemDen}.
The mean $T_\mathrm{e}$\,[N\,{\sc ii}] and $n_\mathrm{e}$\,[S\,{\sc ii}] over the whole nebula are $\sim$9600~K and $\sim$380~cm$^{-3}$, respectively. 
The uncertainty or standard deviation on the density distribution is relatively large (especially in the region a2), but we note that the values derived in different areas are generally consistent. 
The He~{\sc ii} $\lambda$4686 \AA\ emission line is detected with an 
He~{\sc ii} to H$\beta$ ratio ranging from 0.37$\pm$0.04 to 0.97$\pm$0.21, 
with a mean value of 0.61$\pm$0.04 that indicates a relatively excited 
nebula. The mean ionic and total abundances obtained for regions where values of $T_\mathrm{e}$ and $n_\mathrm{e}$ 
could be derived are presented in Table~\ref{Ion}. For this purpose the {\sc PyNeb} 
atomic data and \citet{DIMS14}'s ionic correction factors (ICFs) were used. \\

The emission detected in the GTC slits a and b extends 58$\farcs$4 and 48$\farcs$5,  
respectively, i.e., no emission beyond the extent of the nebula disclosed in the 
narrow-band images is detected (see Fig.~\ref{NOT}).   
However, a careful background subtraction of the GTC slit c unveils the presence of 
two faint [N\,{\sc ii}] blobs, one located 20$\farcs$3 south from position c2 and 
noted c1 in Figure~\ref{slit}, and another one at 42$\farcs$4 north from position 
c4 and noted c6 in this same figure  (see also Fig.\ref{profiles}--bottom panel).    
These emissions suggest the presence of the relics of an external structure  
along a direction orthogonal to that of the main nebula (see below).\\

Figure~\ref{profiles} shows the spatial profiles of various emission lines across 
slit c.  
It is worth highlighting the asymmetry in terms of fluxes, with H$\alpha$, 
[O\,{\sc iii}], and [N\,{\sc ii}] emissions being brighter in the northern 
arm than in the southern one.  
The flux difference is less noticeable for [S\,{\sc ii}] and [Ar\,{\sc iii}]. 
Note also that the high excitation lines, [O\,{\sc iii}] and [Ar\,{\sc iii}] 
(as well as H$\alpha$), are spatially concentrated towards the interior of 
the nebula, contrary to the low excitation lines [N\,{\sc ii}] and [S\,{\sc ii}]. 
Whereas some of these features could be due to the exact positioning 
of the slit, as the images do not show a brightness difference between 
both filaments, a noticeable ionization stratification is present within 
Sab\,99.

\section[]{Infrared emission}\label{archive}

\begin{figure}
\begin{center}
\includegraphics[width=\linewidth]{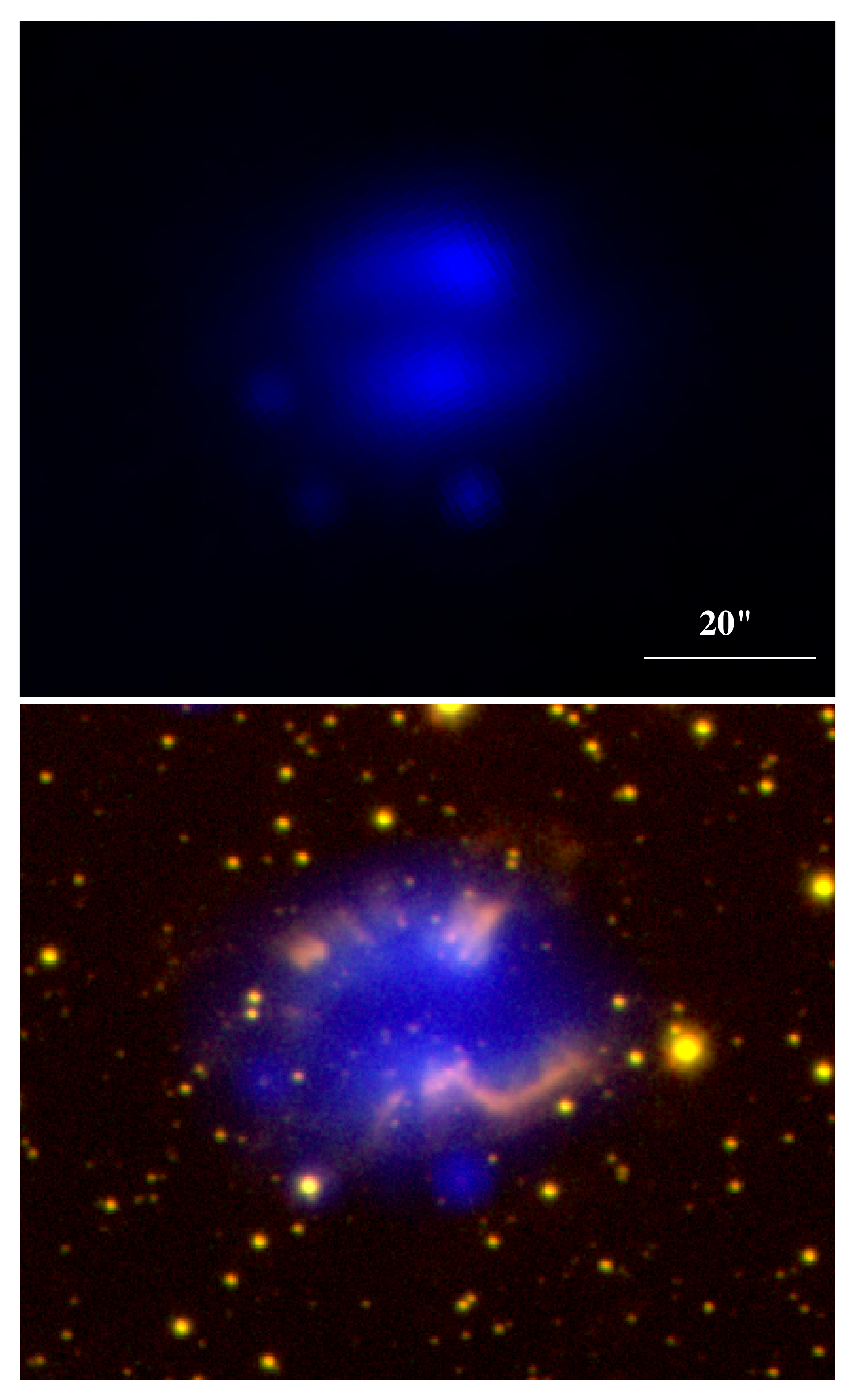}
\caption{
(top) \emph{Spitzer} MIPS 24 $\mu$m band image of Sab\,99.  
The emission shows two hemispheres separated by a dark lane. 
(bottom) NOT ALFOSC and \emph{Spitzer} MIPS 24 $\mu$m color composite 
picture of Sab\,99 with red, green, and blue corresponding to \,[N\,{\sc ii}], 
H$\alpha$ and the 24\,$\mu$m emission.  
The two optical arms encompass very well the mid-IR emission, while the dark 
lane is coincident with the optical innermost regions of the nebula. 
North is top, east to the left.
}
\label{MIPS}
\end{center}
\end{figure}

The mid-IR emission at 24~$\mu$m revealed by the \emph{Spitzer} MIPS image in 
Figure~\ref{MIPS} is spatially consistent with the optical emission; the mid-IR 
emission shows two opposite arms, whose curvature mimics that of the optical 
ones, separated by a dark lane.  
The two brightest mid-IR clumps are coincident with the two brightest 
optical knots denoted as L-M and R-S in Figure~\ref{Kine2}.  
It is interesting to note that the infrared emission is fully enclosed 
within the optical boundaries of the PN ( bottom panel of Fig.~\ref{MIPS}, 
the three IR spots located outside the PN area are actually associated 
with background stars).

This spatial distribution suggests that the mid-IR emission in the 24~$\mu$m 
MIPS band could be attributed to the high-excitation [Ne\,{\sc v}]
$\lambda$24.3 $\mu$m and [O\,{\sc iv}] $\lambda$25.9 $\mu$m emission lines rather than to cold dust \citep{Chu2009}, but there are no available IR spectra to confirm it.

\begin{table*}
\begin{center}
\caption{\label{TemDen}  Determination of the electronic temperatures and densities for the brightest knots using the Monte Carlo procedure.}
\begin{tabular}{@{\extracolsep{4pt}}l|c|c|c|c|c|c|c|c|@{}}
\hline
\hline
     & \multicolumn{2}{c}{Region a2} & \multicolumn{2}{c}{Region b3}& \multicolumn{2}{c}{Region b4} & \multicolumn{2}{c}{Region c4}\\
     & Mean  & $\sigma$ & Mean  & $\sigma$ & Mean  & $\sigma$ & Mean  & $\sigma$ \\
\cline{2-3} \cline{4-5} \cline{6-7}  \cline{8-9}
\hline
$T_\mathrm{e}$\,[N\,{\sc ii}](K) & 9570 & 990 &   9090 & 920 & 10850 &    1090 &8790 &     650\\
$n_\mathrm{e}$\,[S\,{\sc ii}](cm$^{-3}$) &  280 & 240 &  465  & 350 & 410 &     330 & 350    &  230\\
\hline
\hline
\end{tabular} 
\end{center}
\end{table*}

\begin{table*}
\begin{center}
\caption{\label{Ion}  Mean ionic and total abundances in logarithmic scale for the individual brightest knots in the upper and lower arms using a Monte Carlo Procedure.}
\begin{tabular}{@{\extracolsep{4pt}}l|c|c|c|c|c|c|c|c|@{}}
\hline
Ion & \multicolumn{2}{c}{Region a2} & \multicolumn{2}{c}{Region b3}& \multicolumn{2}{c}{Region b4} & \multicolumn{2}{c}{Region c4}\\
    &  N(X)/N(H+) & $\sigma$  &   N(X)/N(H+)   &  $\sigma$ &   N(X)/N(H+)   & $\sigma$  &   N(X)/N(H+)   & $\sigma$ \\
\cline{2-3} \cline{4-5} \cline{6-7}  \cline{8-9}
\hline
   He$^{1+}$&  $-$1.01  &    0.08& $-$1.26  &    0.10  &  $-$1.22  &   0.10 &  $-$1.10  &   0.07   \\
   He$^{2+}$&  $-$1.00  &    0.13& $-$1.20  &    0.14  &  $-$1.27  &   0.16 &  $-$1.09  &   0.10   \\
   N$^{0}$  &  $-$4.83  &    0.27&  ...     & ...      & ...       & ...    &   ...     & ....     \\
   N$^{1+}$ &  $-$3.90  &    0.16& $-$4.04  &    0.18  &  $-$4.28  &   0.17 &   $-$4.12 &    0.13 \\
   O$^{0}$  &  $-$4.15  &    0.20& $-$4.18  &    0.21  &  $-$4.57  &   0.20 &   $-$4.49 &   0.16     \\
   O$^{1+}$ &  $-$3.41  &    0.34& $-$3.35  &    0.36  &  $-$3.83  &   0.32 &   $-$3.49 &    0.27 \\
   O$^{2+}$ &  $-$3.36  &    0.34& $-$3.35  &    0.36  &  $-$3.76  &   0.32 &   $-$3.60 &    0.29  \\
   S$^{1+}$ &  $-$5.59  &    0.16& $-$5.59  &    0.16  &  $-$5.85  &   0.16 &   $-$5.62 &    0.12  \\
   S$^{2+}$ &  ...      & ...    & ...      & ...      & ...       & ...    &   $-$4.87 &    0.18\\
   Ar$^{2+}$&  $-$5.64  &    0.16& $-$5.61  &     0.17 &  $-$5.79  &   0.17 &   $-$5.51 &    0.13      \\
   Ne$^{2+}$& ...       & ...    & ...      & ...      & ...       & ...    &   $-$4.12 &    0.13 \\
 \hline 
He/H       &   $-$0.70  &   0.10 &  $-$0.93 &    0.12  &  $-$0.94  &  0.12  &  $-$0.79  &  0.08\\
O/H        &   $-$2.88  &   0.35 &  $-$2.82 &    0.36  &  $-$3.31  &  0.33  &  $-$3.04  &  0.28        \\
N/H        &   $-$3.42  &   0.18 &  $-$3.56 &    0.18  &  $-$3.81  &  0.18  &  $-$3.71  &  0.14    \\
Ar/H       &   $-$5.42  &   0.18 &  $-$5.37 &    0.18  &  $-$5.58  &  0.18  &  $-$5.30  &  0.14  \\
S/H        &   $-$5.01  &   0.16 &  $-$5.03 &    0.17  &  $-$5.26  &  0.16  &  $-$5.12  &  0.13 \\
Ne/H       &    ...     &  ...   &  ...     &   ...    &    ...    & ...    &  $-$3.88  &  0.15     \\
\hline
\end{tabular} 
\end{center}
\end{table*}

\section[]{Morpho-kinematic modelling}\label{model}

Using the morphological and kinematical information derived from the high resolution 
NOT images and high-dispersion MES observations, it is possible to build a spatio-kinematic 
model of Sab\,99 to determine its physical structure. 
For this purpose we will use the spatio-kinematic code {\sc shape} \citep{Steffen2011}.

The PV diagrams obtained from slits A and B across the northern arm indicate the 
presence of a number of knots aligned along a blue-shifted curve, whereas those 
of slits D and E shows a number of mostly red-shifted knots also along a curve 
(Fig.~\ref{Kine1}). 
The position and velocity of these knots are indicative of a tilted torus.  
Combined with the remaining spectra associated to the slits passing through both 
arms, i.e. slits F, G, and H, we notice some degree of symmetry between opposite 
knots, suggesting that they are all indeed contained in a truncated torus that 
contains regions of varying density.

\begin{figure*}
\begin{center}
\includegraphics[width=0.8\linewidth]{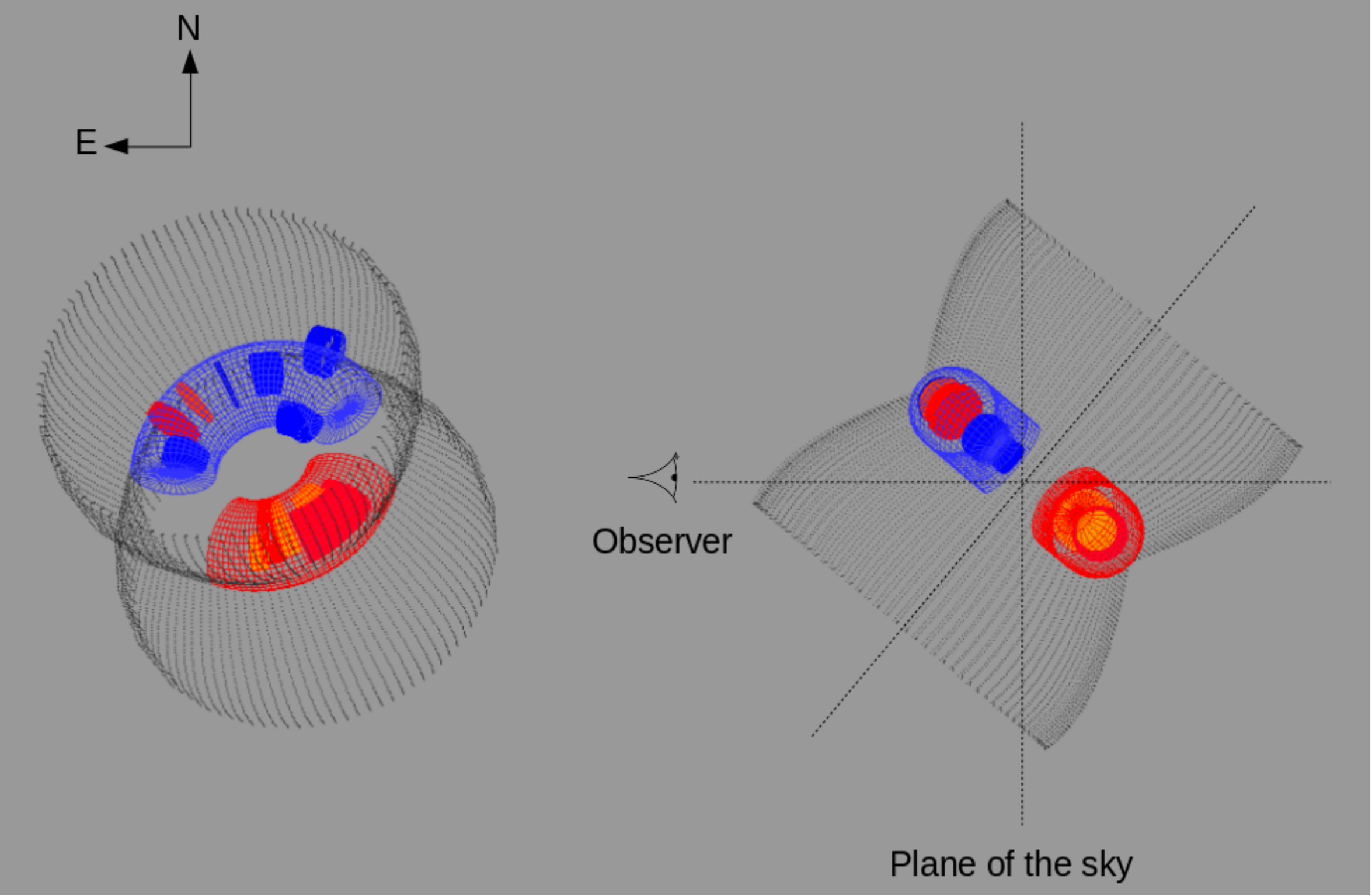}
\caption{
Schematic images of Sab\,99 based on our best fit spatio-kinematic {\sc shape} model 
projected onto the plane of the sky (left) and rotated by 90$^\circ$ (right). 
The color code for the knots indicates the blue-- and red-shifted components, where 
an additional orange color in the red-shifted region has been used to distinguish 
among adjacent knots. 
The length and morphology of the bipolar outflows are only presented for 
illustrative purposes.
}
\label{model1}
\end{center}
\end{figure*}

\begin{figure*}
\begin{center}
\includegraphics[width=1\linewidth]{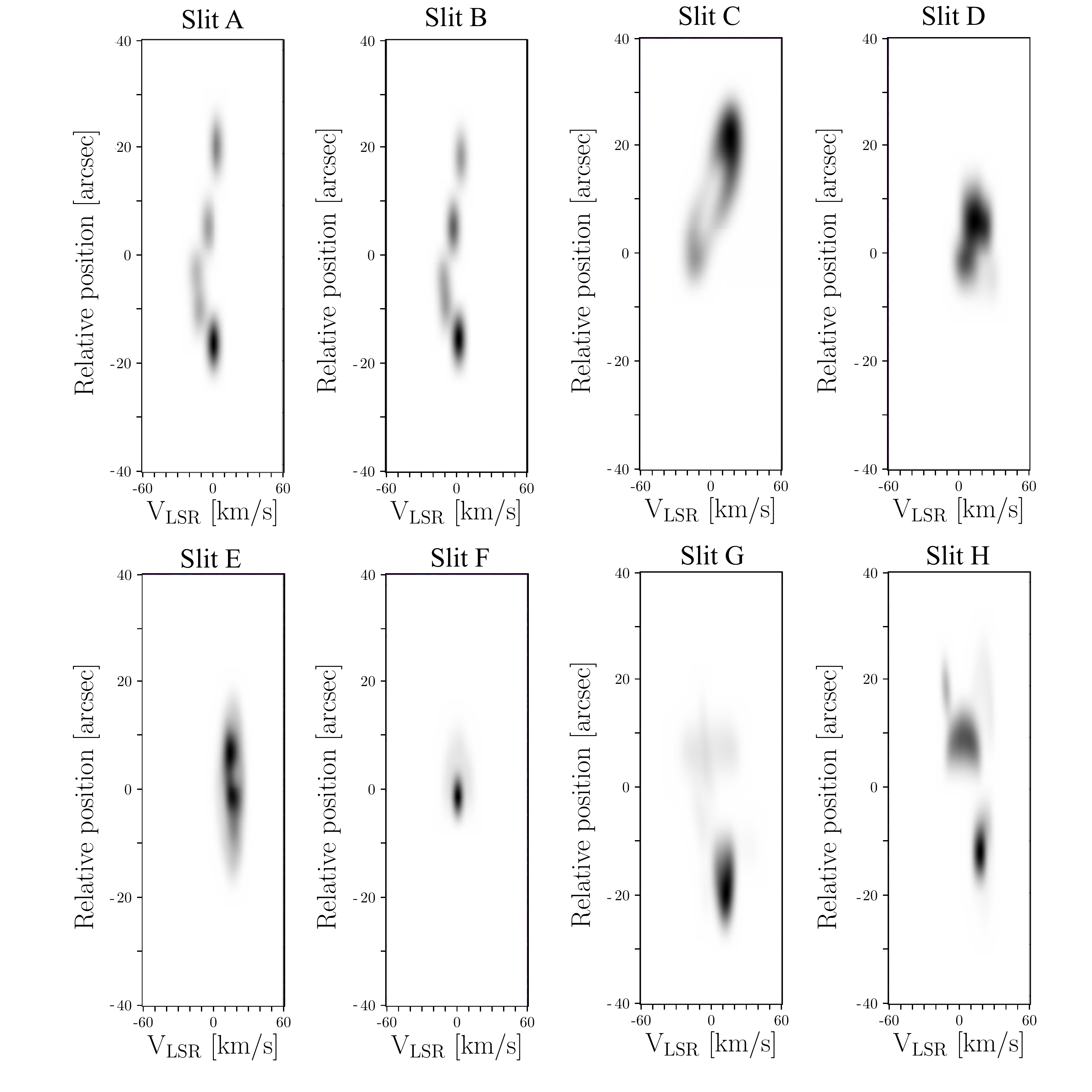}
\vspace*{-0.8cm}
\caption{
Synthetic PV diagrams of the echellograms shown Figure~\ref{Kine1} obtained from 
the best fit spatio-kinematic {\sc shape} model shown in Figure~\ref{model1}.  
}
\label{PVsynt}
\end{center}
\end{figure*}

The spatio-kinematic model of Sab\,99 is presented in Figure~\ref{model1}.  
The nebula is mostly composed of a broken torus $\simeq$40\arcsec\ in 
length whose symmetry axis is tilted by $\simeq$40$^\circ$ with the 
plane of the sky. 
The systemic velocity of the torus would be $v_{\rm LSR} \approx 10$ km~s$^{-1}$.  
The synthetic PV diagrams produced at the location of the MES slits 
in Figure~\ref{PVsynt} provide a satisfactory fit to the observed PV 
diagrams (Fig.~\ref{Kine1}), with a number of blue- and red-shifted 
knots.  
We note that the faint surface brightness and fragmented appearance of Sab\,99 
hamper an accurate determination of the inclination, size, and systemic velocity 
of this torus.  
Different models with the knots being relatively co-planar within an inclination 
with the plane of the sky of the torus axis in the range 35$^{\circ}$--45$^{\circ}$ 
produce synthetic PV diagrams consistent with those observed. 
The large variance of the radial velocities of these knots, mostly in the 
northern arm, suggests a gradual uncoupling from the original structure.

The high-dispersion MES observations did not detect the emission from the 
structure registered in the much deeper GTC OSIRIS observations along the 
slit "c" as marked by the labels "c1" and "c6" in Figure~\ref{slit}. 
Such information is not sufficient to constrain the shape and extent of 
this structure, which we propose is the pair of bipolar lobes orthogonal 
to the disrupted torus of Sab\,99 illustrated in Figure~\ref{model1}.

\section[]{Discussion}\label{summary}

\subsection{The true shape of  Sab\,99}

The morphology of Sab\,99 could be associated  {\it at first} to the broken edges of an evolved bipolar PN 
 with a pinched waist. The bipolar axis of such a PN seems oriented along the NW to SE direction with PA $\sim-$65$^{\circ}$.
Sab\,99 would be morphologically similar to PNe with pinched waists, 
such as NGC\,2818 \citep{Vazquez2012}, NGC\,650-1 \citep{Ramos2018} 
and PN\,G321.6+02.2 \citep{Corradi1997}, but the lack of kinematic evidence for a fast outflow along PA $-65^{\circ}$ seems to question this interpretation. 


The deep GTC OSIRIS spectra, however, unveil the presence of two faint clumps 
located on either side of the main structure along a 
direction at PA $\sim-$10$^{\circ}$.
Thus,  Sab\,99 would actually be a fragmented ring-like structure with the faint remains 
of a bipolar outflow whose axis is oriented along the NE to SW direction.
 The detailed orientation and geometry of this outflow and the ring-like structure have 
been constrained with the kinematic and morphological information using a {\sc shape} 
model (see previous section).
In this case, Sab\,99 would rather compare to objects such as NGC\,5189 
\citep{Sabin2012a} or NGC\,6309 \citep{Vazquez2008}. 
To some extent we can also consider the morphological similarities with 
IPHASX J194359.5+170901 \citep[ or PN G054.2-03.4, a.k.a.\ the Necklace Nebula,][]{Corradi2011}, 
with its well defined ring of knots and remnant of bipolar ejecta. 


\subsection{A distant and old PN}

The distance is a fundamental physical parameters in the study of a PN, 
but it is also one of the least well determined. 
The broken morphology of Sab\,99 makes uncertain any determination of an average 
radius, which is a critical datum to determine the distance to a PN using the 
statistical method based on the relationship between the H$\alpha$ surface brightness ($S_\mathrm{H\alpha}$) and the radius of a PN \citep{Frew2016}. 
Alternatively, the distance can be derived from the extinction to the PN 
if the relationship between the extinction and the distance along the line 
of sight is known.  

Since the extinction derived at different locations of the nebula is relatively constant, we can assume that the circumstellar extinction is negligible. 

\begin{figure}
\begin{center}
\includegraphics[width=\linewidth]{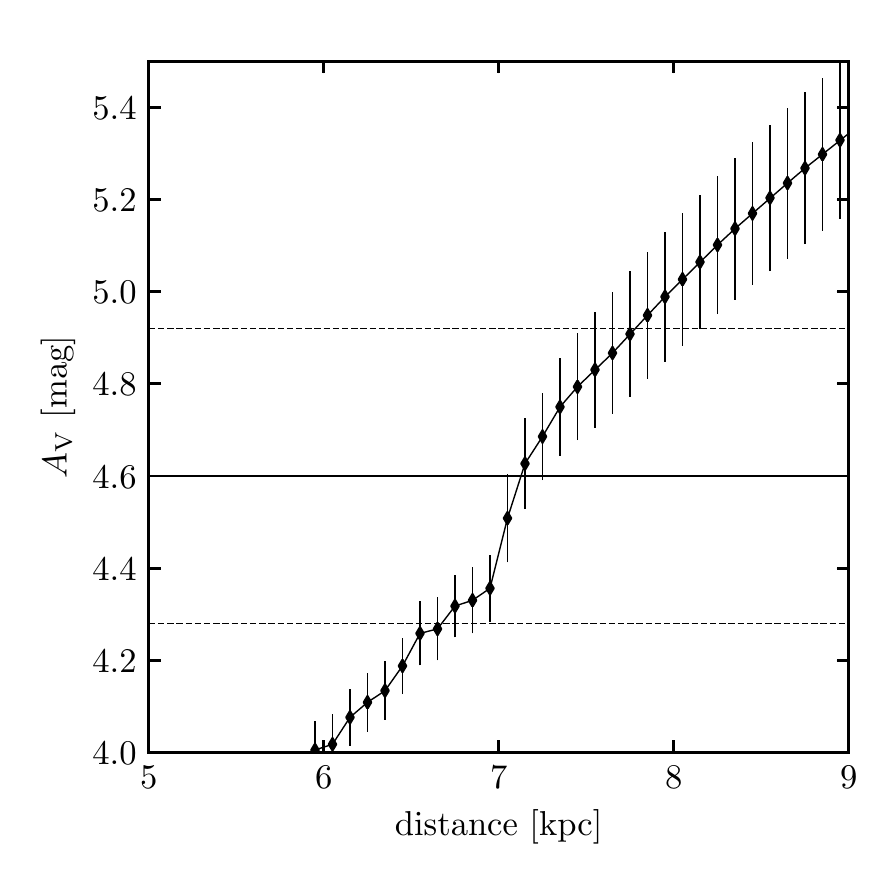}
\vspace*{-0.8cm}
\caption{Distance-extinction relationship derived from a extinction map  close to the line of sight of Sab\,99. Only the portion of the curve with the largest slope where it is possible to derive distance is shown. The extinction of Sab\,99 is denoted by a horizontal solid line at A$_{V}$ = 4.6 mag, with the horizontal dotted lines marking the 1$\sigma$ uncertainties.  
}
\label{dist}
\end{center}
\end{figure}

Detailed 3D extinction maps for the Northern Galactic plane have been generated 
using photometric data obtained in the framework of IPHAS \citep{Sale2014}.
This extinction-distance method was applied by \citet{Giammanco2011} 
to a sample of known PNe using earlier versions of those extinction 
maps and concluded that distances could be obtained with an accuracy better 
than 35\% depending on the uncertainty on the determination of the 
nebular extinction and on the quality of the extinction map along this 
exact line of sight.  The authors also described the main sources of error associated with the method.  
As for Sab\,99, we will adopt the 3D extinction map centered at 
$l = 056.125^{\circ}$, $b = -00.542^{\circ}$ (Fig.~\ref{dist}). 
The mean c(H$\beta$) value of 2.14$\pm$0.15 derived in previous sections 
implies an extinction $A_\mathrm{V}$ = 4.6$\pm$0.3 that intersects the extinction 
to distance curve in Figure~\ref{dist} at a distance of 7.1$_{-0.3}^{+0.8}$ 
kpc. 

The main caveat in this distance estimation is that the source is centered at $\simeq$7\farcm8 from the center of the $10^\prime\times10^\prime$ map and the variation on the extinction might play an important role. 
We examined other adjacent maps \footnote{http://www.iphas.org/extinction/} and found the closest one with trustworthy results on the distance for the value of  $A_\mathrm{V}$ previously derived to be centered at $l = 056.125^{\circ}$, $b = -00.375^{\circ}$.  
It indicates a distance of 6.3$_{-0.9}^{+0.5}$~kpc for Sab\,99, thus the distance of 7.1~kpc can be considered as an upper limit taking into account variations on the visual extinction along the line of sight.

The velocity distribution indicates the overall motion of the central nebula with the 
northern section of the ring being mostly blue-shifted and 
the southern counterpart being red-shifted. 
The values of the radial velocities are moderate, with a maximum of 30.1 km~s$^{-1}$. 
Taking into account the geometry obtained with our {\sc shape} model, where the central 
filamentary structure is an expanding torus following an homologous law, a mean expansion 
velocity of 25$\pm$5 km~s$^{-1}$ and a mean radius of 19\arcsec$\pm$10\arcsec can be 
estimated.  

At a distance of 7.1 kpc, this results in a radius of 0.65$^{+0.42}_{-0.37}$ pc 
and a kinematical age $\lesssim$26000 yr.
 
Although the latter value is highly affected by the large uncertainties in distance, ring radius and expansion velocity, and depends on the strong assumption of an 
expansion at constant speed, a relatively late evolutionary stage is consistent with the fragmented and diffuse appearance of Sab\,99.

\subsection{Estimated physical and chemical characteristics}

 The high excitation [O\,{\sc iii}] $\lambda$4363\AA, [Cl\,{\sc iii}] or [Ar\,{\sc iv}] 
emission lines are absent from our spectra and most of the calculations referring to the 
physical parameters and abundances were performed using low excitation ions. 
The physical analysis reveals a relatively homogeneous nebular temperature (within the error bars) with <$T_\mathrm{e}$\,[N\,{\sc ii}]> $\approx$ 9600~K and low density <$n_\mathrm{e}$\,[S\,{\sc ii}]> $\approx$ 380 cm$^{-3}$. 
A comparison with the much younger PN NGC\,6309, with structures aged between 3700 and 
4000 years, indicates roughly similar temperatures ($T_\mathrm{e}$\,[N\,{\sc ii}] = 10200--11800 K), but much higher densities ($n_\mathrm{e}$\,[S\,{\sc ii}] = 1600--3800 cm$^{-3}$) in the central 
knotty torus \citep{Vazquez2008}.  
The brightest knots in Sab\,99 are particularly rich in He, N and O (Table \ref{Ion}). The average <N/O> value of $\approx$ 0.235 and average <He/H> value of $\approx$ 0.148 can be used to establish the nature of the PN. Hence, based on these abundance ratios and \citet{Kingsburgh1994}'s definition, Sab\,99 is a non-Type I PNe as N/O$\leq$0.8 although we note that the He/H abundance is large and more typical of a Type I. In addition, the \citet{Peimbert1978} classification scheme identifies the PN as a Type II based on the same N/O ratio.
Sab\,99 can be considered as a relatively high excitation nebula as indicated by the presence of the He\,{\sc ii} $\lambda$4686\AA\ emission line (with I/H$\beta$ up to 105$\pm$25 in the Region c and averaging 65$\pm$12 in the whole nebula) and a mostly oxygen-rich nebula. 

Therefore, Sab\,99 is likely to have an intermediate mass in the $\approx$1.2--2.0~M$_\odot$ range. The evolutionary models of \citet{Marigo2001} can be used to provide some constraints 
on the progenitor initial mass (M$_\mathrm{i}$) based on the mean He/H, log~(N/H) and log~(O/H) of Sab\,99. This results in an initial mass estimate $\approx$2--3~M$_{\sun}$ for a metallicity 0.019~Z$_\odot$ i.e. the upper range limit for this intermediate mass PN. Following the method by \citet{Maciel2010} using the N/O abundance ratio, the central star mass ($M_\mathrm{CSPN}$) is estimated to be $\sim$ 0.61 M$_{\sun}$.


We conducted a simple {\sc Cloudy} \citep{Ferland2017} photoionisation modelling for a nebula with the physical structure of the central ring described by our {\sc shape} model and the chemical abundances as those listed in Table~\ref{Ion}. A model atmosphere by \citet{Rauch2003} was adopted to describe the emission from the central star. The best fit models for the low and high excitation emission lines imply an effective temperature log($T_\mathrm{eff}$) $\approx$ 5--5.2 and luminosities in the range log($L$/L$_{\sun}$) $\approx$ 3.5--4. From the models, and assuming the reddening $A_\mathrm{V}$ = 4.6$\pm$0.3, we estimated that the visual magnitude of the central star is fainter that $\sim$21--22 mag. This is in concordance with the magnitudes of the field's stars detected in the central area of Sab\,99 showing {\it r'} mag fainter than $\sim$20 \citep{Barentsen2014}. Again, the errors on the different parameters involved have to be taken into account, but the results tend to support the fact that the central star is not visible.


The high effective temperature of the central star of Sab\,99 and the detailed spatial distribution of the emission in the \emph{Spitzer} MIPS 24~$\mu$m band  (Fig.~\ref{MIPS}) provides additional support to the origin of this emission from the [O\,{\sc iv}] line at 25.9~$\mu$m \citep{Chu2009}. 

We would therefore be detecting emission lines from highly ionized material 
inside the spatial boundaries of the optical emission and indeed the {\sc Cloudy} 
model described above implies a flux for this line 8 times that of the H$\beta$ line. 
Whereas dust may still be present inside the nebula, we note that there are no 
variations (within the errors) of c(H$\beta$) across the nebula, suggesting 
a homogeneous dust distribution.
Mid-IR spectroscopic observations are required to fully settle down this issue.

\section[]{Conclusions}\label{Con}

We have presented a detailed study of the spatial, kinematic, spectroscopic, 
and chemical properties of the newly discovered IPHAS PN Sab\,99 based on 
sub-arcsec optical narrow-band images and high-dispersion echelle and deep 
optical intermediate-dispersion spectra.  
This investigation has allowed us to achieve sound conclusions on different 
aspects of this source.

The physical structure of Sab\,99 is bipolar, but, contrary to naive expectations 
based on its optical morphology, the two opposite arm-like structures are 
fragmented remains of an equatorial torus aligned along the East-West direction, 
with an extremely faint bipolar outflow along the North-South direction detected 
in deep GTC OSIRIS spectra that escapes detection in our NOT ALFOSC optical 
narrow-band images.  
Deeper images in the [N~{\sc ii}] emission could provide information on its 
detailed morphology and extent.  
The kinematics of individual knots in this torus indicates 
their gradual uncoupling and destruction of this structure.

The progenitor star of Sab\,99 had an initial mass of $\approx$2--3~M$_{\sun}$, as implied by the comparison of the chemical abundances of the nebula, typical of a Type~II PN, with the expectations of models of chemical evolution of low- and intermediate-mass stars.  
The mass of the stellar remnant is also significant, $\sim$0.61 M$_{\sun}$. 
According to simple {\sc Cloudy} photoionization models, the central star has nowadays a high effective temperature, around 10$^{5}$~K, resulting in notable nebular emission in the high excitation lines of He\,{\sc ii} $\lambda$4686 \AA\ and perhaps [O\,{\sc iv}] 25.9 $\mu$m.

 The nebula is assumed at a distance of 7.1$_{-0.3}^{+0.8}$ kpc, obtained with the IPHAS extinction maps and owing the variations of interstellar extinction in the line of sight. This in turns implies a kinematic age $\lesssim$26000 yr. 
Although this figure is affected by large uncertainties, a late evolutionary stage 
for Sab\,99 would be consistent with the low density and disrupted morphology and 
kinematics of its equatorial ring, as well as the high effective temperature of 
its central star.

Although Sab\,99 is inherently faint and complex, our investigation has been able to conclude that it was once a bipolar PN of which only a disrupted equatorial ring remains, while the bipolar lobes are no longer visible, except for some remaining patches.  
The torus at the nebular equator is still present but shows spatial 
and kinematic signs of a gradual dismantlement. 
Its central star has a high effective temperature.  
Such evolved object is one kind of PNe that can be expected to be found in IPHAS.

\section*{Acknowledgments}

 The authors are thankful to the referees for their comments that improved the present paper.
LS acknowledges support from PAPIIT grant IN-101819 (Mexico).  
MAG acknowledges support from grant AYA PGC2018-102184-B-I00
co-funded with FEDER funds. SZ works under the collaboration agreement "UNAM-TecNM
43310-3020-30-IX-15". JAT and MAG are funded by UNAM DGAPA PAPIIT project IA100720. 
GRL acknowledges support from CONACyT (grant 263373) and PRODEP (Mexico).

This work is partially based on observations collected at the Observatorio Astron\'omico Nacional at San Pedro M\'artir, B.C., Mexico and we thank the daytime and night support staff at the OAN-SPM for facilitating and helping obtain our observations.
Based on observations made with the Nordic Optical Telescope, operated by the Nordic 
Optical Telescope Scientific Association, and the Gran Telescopio Canarias (GTC), 
both installed in the Spanish Observatorio del Roque de los Muchachos of the Instituto 
de Astrof\'\i sica de Canarias in La Palma, Spain. 
This paper makes use of data obtained as part of the INT Photometric H$\alpha$ Survey of the Northern Galactic Plane (IPHAS, www.iphas.org) carried out at the Isaac Newton Telescope (INT). The INT is operated on the island of La Palma by the Isaac Newton Group in the Spanish Observatorio del Roque de los Muchachos of the Instituto de Astrofisica de Canarias. All IPHAS data are processed by the Cambridge Astronomical Survey Unit, at the Institute of Astronomy in Cambridge. The bandmerged DR2 catalogue was assembled at the Centre for Astrophysics Research, University of Hertfordshire, supported by STFC grant ST/J001333/1.
The OSN director is acknowledged for awarding observations through a DDT program 
and the telescope operator Alfredo Sota for conducting the observations.
This work is based in part on observations made with the {\it Spitzer} Space Telescope, 
which is operated by the Jet Propulsion Laboratory, California Institute of Technology under a contract with NASA.
In memoriam of Johannes Andersen, director of the NOT from 2002 to 2013.  

\section*{Data availability}

The data underlying this article will be shared on reasonable request to the corresponding author.

\bibliographystyle{mn2e}

\bibliography{sabin}

\bsp

\label{lastpage}

\end{document}